\documentclass[twocolumn,showpacs,aps]{revtex4}

\usepackage{graphicx}
\usepackage{dcolumn}
\usepackage{bm}

\begin{document}
\title {Comments on "Vortex phase diagram of HgBa$_2$Ca$_2$Cu$_3$O$_{8+\delta}$ thin films from magnetoresistance measurements"}
\author {Y. Z. Zhang}
\author{Z. Wang}
\affiliation{National Laboratory for superconductivity, Institute
of Physics \& Center for Condensed Matter Physics, Chinese Academy
of Sciences, P. O. Box 603,  100080, Beijing, China}

\begin{abstract}
We make comments on the paper  presented by Kim \textit{et al.}
[Phys. Rev. B \textbf{61}, 11317 (2000)]. The authors analyzed  activation energies of
HgBa$_2$Ca$_2$Cu$_3$O$_{8+\delta}$ thin films  with a scaling
relation, and defined four vortex regions for thin films. We
find that the definitions of  four vortex regions  and  the
scaling relation are questionable when studying their definition
of the resistivity range for thermally activated flux flow. Using
the empirical activation energy form suggested by Zhang \textit{et
al.}  [Phys. Rev. B \textbf{71},  052502 (2005)] for lower resistivity data, we find that the
form is not only in good agreement with the resistivity
$\rho(T,H)$, but also with the apparent activation energy
$-\partial \ln \rho(T,H)/\partial T^{-1}$.
\end{abstract}

\maketitle Previously, Kim \textit{\textit{et al.}} \cite{Kim}
reported an investigation of resistive behaviors of
HgBa$_2$Ca$_2$Cu$_3$O$_{8+\delta}$ (Hg-1223) thin films in the
mixed state. A scaling relation was proposed for analyzing the
resistive behavior and was found in good agreement with the
apparent activation energy  $-\partial \ln \rho(T,H)/\partial
T^{-1}$ (the so-called effective activation energy in this paper)
in a specialized resistivity range.  With this scaling, they
suggested that the vortex system could be divided into four
different vortex regions corresponding to the flux flow (FF)
region, the thermally activated flux flow (TAFF) region,  the
critical  state region, and the vortex solid (VS) region.  Three
characteristic temperatures corresponding to $T_{ff}$, $T^*$, and
$T_{irr}$ were defined for the boundaries of these regions. They
claimed that the TAFF behavior was limited in the region of
$T^*<T<T_{ff}$ and the corresponding activation energy was
expressed as $U_0(T,H) \sim H^{-1.1}(1-T/T_c)^{1.5}$ for the
magnetic field range from 1 to 9 T. In this comment, we point out
that the scaling relation and the definitions of the four
different vortex regions are questionable. We propose that the
activation energies of Hg-1223 thin films relate to lower resistivity where $T<T^*$. Using
lower resistivity data, we find that the activation energy  is
expressed as the form suggested by Zhang \textit{et al.} \cite
{Zhang}.

After the discovery of  high temperature superconductors
(HTSCs),  activation energies of different HTSCs have been
widely studied in theories and experiments.  According to Palstra
\textit{et al.} \cite{Palstra}, activation energies of HTSCs
should be determined in the temperature interval over which the
resistivity changes from $10^{-4}$ to $1$ $\mu\Omega$ cm or for
the resistivity ratio $\rho(T,H)/\rho_n$ approximately below
$1\%$, where $\rho_n$ is normal-state resistivity. The temperature
interval is widely accepted for studying  activation energies of
HTSCs; besides, it is widely accepted that TAFF resistivity shall
be ohmic (linear $I$-$V$ relation), and the activation energy is
independent of the applied current density $j$ for $j\to 0$ \cite
{Palstra, Geshkenbein, Vinokur, Blatter, Brandt, Cohen}.  This
means  that  non-ohmic behavior ought to be observed for
decreasing temperature out of  the TAFF region.

\begin{figure}
\includegraphics
[width= .80\columnwidth] {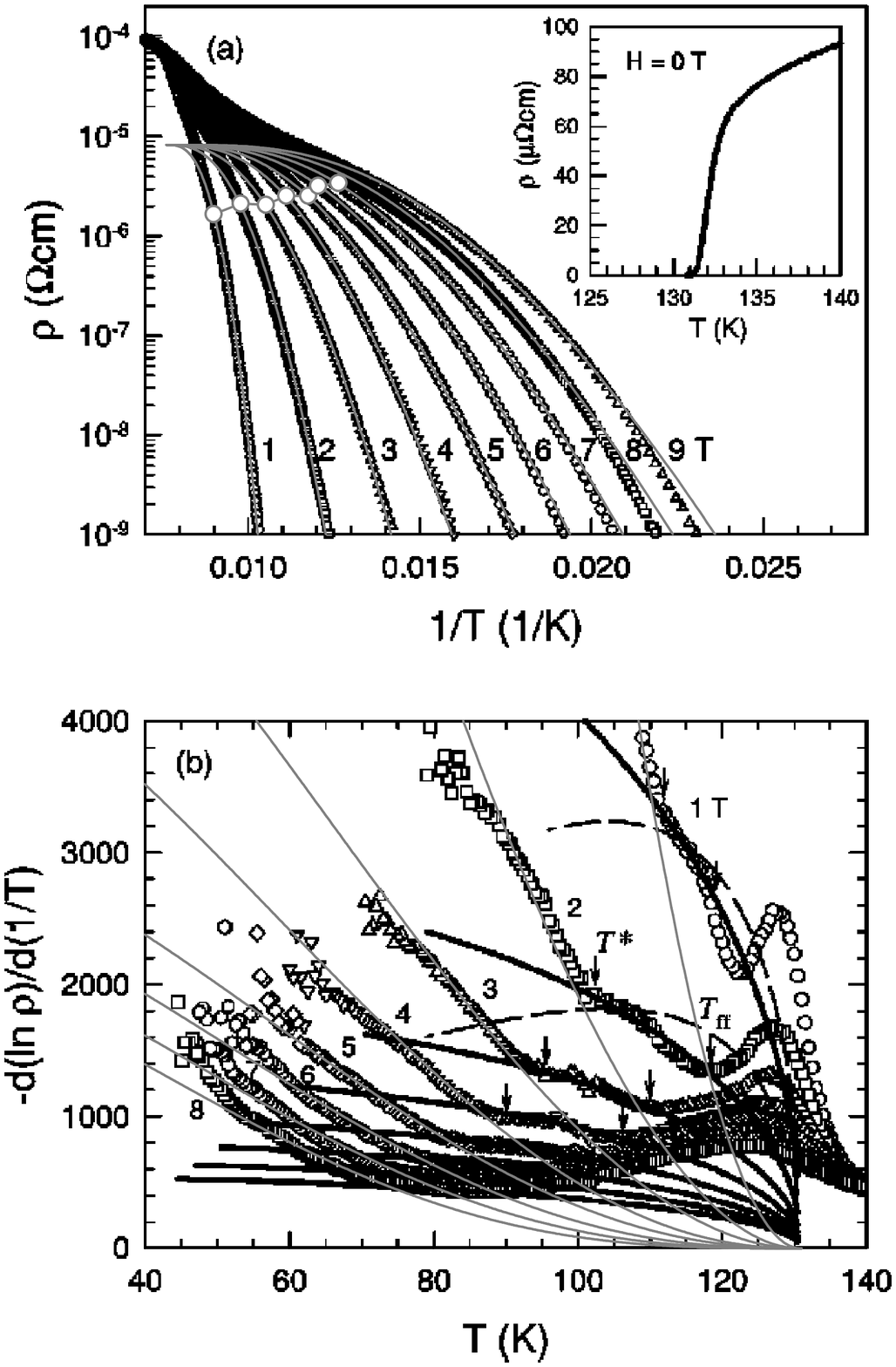} \caption{\label{f1} (a) Arrhenius
plot of $\rho(T,H)$ from Ref.~\cite{Kim}. The gray circles
correspond to $\rho(1/T^*,H)$ data where $H(T^*)$ was determined
from  Fig.4 of Ref.~\cite{Kim}.  (b)$-\partial \ln
\rho(T,H)/\partial T_{ff}^{-1}$ curves from Ref.~\cite{Kim}.  The
black lines,  the black dashed  lines, arrows,  $T^*$, and
$T_{ff}$  are presented  by the authors \cite{Kim}.  The gray
lines in (a) and (b) are our regressions using $\rho_{0f}=8$
$\mu\Omega$ cm for the curves (see text for more details).}
\end{figure}

Using the data of Fig. 1 and Fig. 4 in Ref.~\cite{Kim}, we roughly
redraw  the $\rho (T^*,H)$ data in Arrhenius plot with gray circles as
shown in Figure.~\ref{f1}(a), for which  the data in the range of
$\mu_0H(T^*)\le 7.0$ T were presented due to  $H(T^*)$ data being
only presented for $\mu_0H(T^*)\le 7.0$ T in Fig. 4 of
Ref.~\cite{Kim}.  One will easily find that $T^*(H)$ results in
that TAFF behavior is related to the resistivity value $\rho>1.4$
$\mu \Omega$ cm with resistivity ratio $\rho/\rho_n> 1.5 \%$,
where $\rho_n=\rho(140$ K$, H=0)\approx 93$ $\mu \Omega$ cm. In
this case, Kim \textit{et al.}  gave the TAFF temperature interval
over which the resistivity approximately changes  from $1.4 $ to
$10$ $\mu \Omega$ cm, or  $\rho(T,H)/\rho_n$ approximately above
$1.5\%$. This interval is apparently mismatched with the interval
suggested by Palstra \textit{et al.} Below temperature $T^*$, Kim
\textit{et al.} defined the critical state region, but they did
not further present physical meaning for the region.  In
comparing with other HTSCs, it is very dubious that non-ohmic
resistivity can be found around  the so-called $T^*$.  As a
result,  we argue that  definitions of the TAFF and the
critical state regions in the article are questionable and the
TAFF region ought to relate to the temperature interval as
suggested by Palstra \textit{et al.}; besides, we argue that the
definition of $T_{ff}$ is incorrect in the paper \cite
{Zhang}.

Normally, the TAFF resistivity is expressed as
$\rho=\rho_0\exp(-U_0/T)$ with the activation energy
$U_0(T,H)=U_0(0,H)(1-t)^\beta$, where $t=T/T_c$, $\beta$ is
constant, and $T_c$ is the critical temperature. Considering the
interlayer decoupling in high fields, Zhang \textit{et al.} \cite
{Zhang} suggested  an empirical relation
\begin{equation}
\rho=\rho_{0f}\exp[-U(T, H)/T], \label{eq1}
\end{equation}
with
\begin{equation}
U(T,H)=g(H)f(t), \label{eq2}
\end{equation}
and
\begin{equation}
f=(1-t)^\beta, \label{eq3}
\end{equation}
where $\rho_{0f}$ is constant, and $g$ and $\beta$ are magnetic
field dependent. Using the progression $(1-t)^\beta= 1-\beta
t+\beta(\beta-1)t^2/2!-\beta(\beta-1)(\beta-2)t^3/3!+\ldots$, we
have $\ln \rho \approx (\ln \rho_{0f}+
g\beta/T_c)-(g/T)[1+\beta(\beta-1)t^2/2!-\beta(\beta-1)(\beta-2)t^3/3!+\ldots]$,
where the term $(\ln \rho_{0f}+ g\beta/T_c)\approx \ln \rho_0$ is
temperature independent. With $\beta= 1$, we have $\ln
\rho_0\approx \ln \rho_{0f}+ g/T_c$. However, we find if $\beta$
largely deviates from $\beta= 1$,  the relation of $\ln
\rho_0\approx \ln \rho_{0f}+ g\beta/T_c$ will bring in large
uncertainty and errors for determining $\rho_{0f}$ and $T_c$ in
the relation of  $\ln \rho_0= \ln \rho_{0f}+ g\beta/T_c$ plot (see Ref.\cite
{Zhang}) , as the value of  the local slope $-\partial \ln
\rho(T,H)/\partial T^{-1}$  largely changes from one local
temperature to the other as shown in Fig.~\ref{f1}(b). This means
that we shall determine the $\rho_{0f}$ value in the other way
when $\beta$ largely deviates from $\beta = 1$.

Accordingly,  we must determine four free parameters $T_c$,
$\rho_{0f}$, $g(H)$, and $\beta(H)$ using Eqs.(\ref{eq1}),
(\ref{eq2}), and (\ref{eq3}) with the experimental data.
Generally, $T_c$ and $\rho_{0f}$ are magnetic field independent
and therefore can be eliminated from the free parameter list in
each magnetic field.   In simplicity, we follow the selection of
Kim \textit{et al.} to take $T_c=131$ K (this will not lead to
large errors as the transition width is less than 2 K in  zero
magnetic field and $T_c$ shall be determined around the transition
interval).   We consider the resistivity value for $\rho_{0f}$ (by
trial and error) which is not only getting better regression for
each $\rho(T,H)$  curve in TAFF region, but also is  in good
agreement with each $-\partial \ln \rho(T,H)/\partial T^{-1}$.
Hence, this will leave only two free parameters $g$ and $\beta$
for each magnetic field.  At first, we use the formula
\begin{equation}
U(T,H)\approx T\ln [\rho_{0f}/\rho(T,H)] \label{eq4}
\end{equation}
with $\rho(T,H)$ data to determine the parameters, and then check
the parameters with corresponding regressions for the curves in
Figs.~\ref{f1}(a) and (b).  The gray solid lines in
~Figs.~\ref{f1}(a) and (b) show the results which we take
$\rho_{0f}=8$ $\mu\Omega$ cm for regressions. One will find that
the regressions are in good agreement with the data in the
temperature interval where the resistivity roughly changing from
$10^{-2}$ to $1$ $\mu\Omega$ cm. Note that the interval  matches
the temperature interval suggested by Palstra \textit{et al}.
\cite{Palstra}.

In fact,   a $\rho_{0f}$ value can be selected in a broad range without  changing the 
consistent matches of all the fits for $U(T,H)$  and $\rho(T,H)$ curves in TAFF region. However, changing
$\rho_{0f}$ value will lead to  changes of fitting parameters and
still results in uncertainty in analysis. This means that we can
not  decide  the $\rho_{0f}$ value which is only good for the fits
of $U$ or $\rho$ curves, and we  have to check  the results with
$-\partial \ln \rho(T,H)/\partial T^{-1}$ relation. We find that
the value of $\rho_{0f}$ will  change  the consistent matches of  the fits for
$-\partial \ln \rho(T,H)/\partial T^{-1}$ curves in TAFF region.

\begin{figure}
\includegraphics
[width= .80\columnwidth] {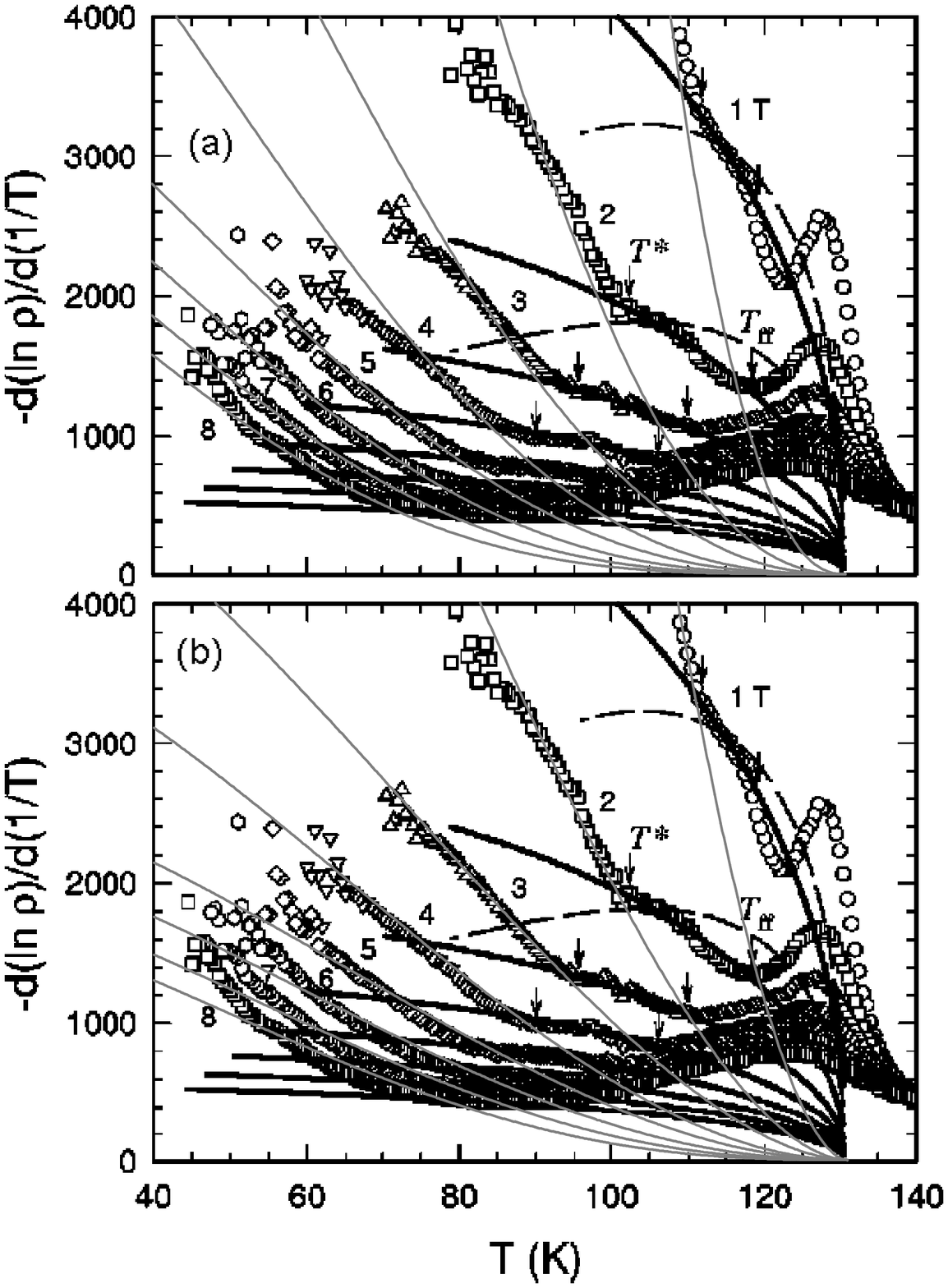} \caption{\label{f2} The gray  lines
in (a) and (b) are our regressions using (a)$\rho_{0f}=6$
$\mu\Omega$ cm, and (b)$\rho_{0f}=10$ $\mu\Omega$ cm for
$-\partial \ln \rho(T,H)/\partial T_{ff}^{-1}$ curves.}
\end{figure}

Figures~\ref{f2}(a) and (b) show the fitting results of $-\partial
\ln \rho(T,H)/\partial T^{-1}$  with $\rho_{0f}=6$ $\mu\Omega$ cm
and 10 $\mu\Omega$ cm, respectively.   The fits are not good in
Fig.~\ref{f2}(a) for low fields and in Fig.~\ref{f2}(b) for high
fields, respectively, while they are in good agreement with
experimental data in  other field ranges. The analysis suggests
that  we have $\rho_{0f}= 8.0\pm 1.5$ $\mu\Omega$ cm for getting
better fits for all   $-\partial \ln \rho(T,H)/\partial T^{-1}$
curves. The deviations between the regressions and 
experimental data in low temperature range are possibly due to
competing relations between coupling and decoupling and between
pinning and depinning.

\begin{figure}
\includegraphics
[width= .80\columnwidth] {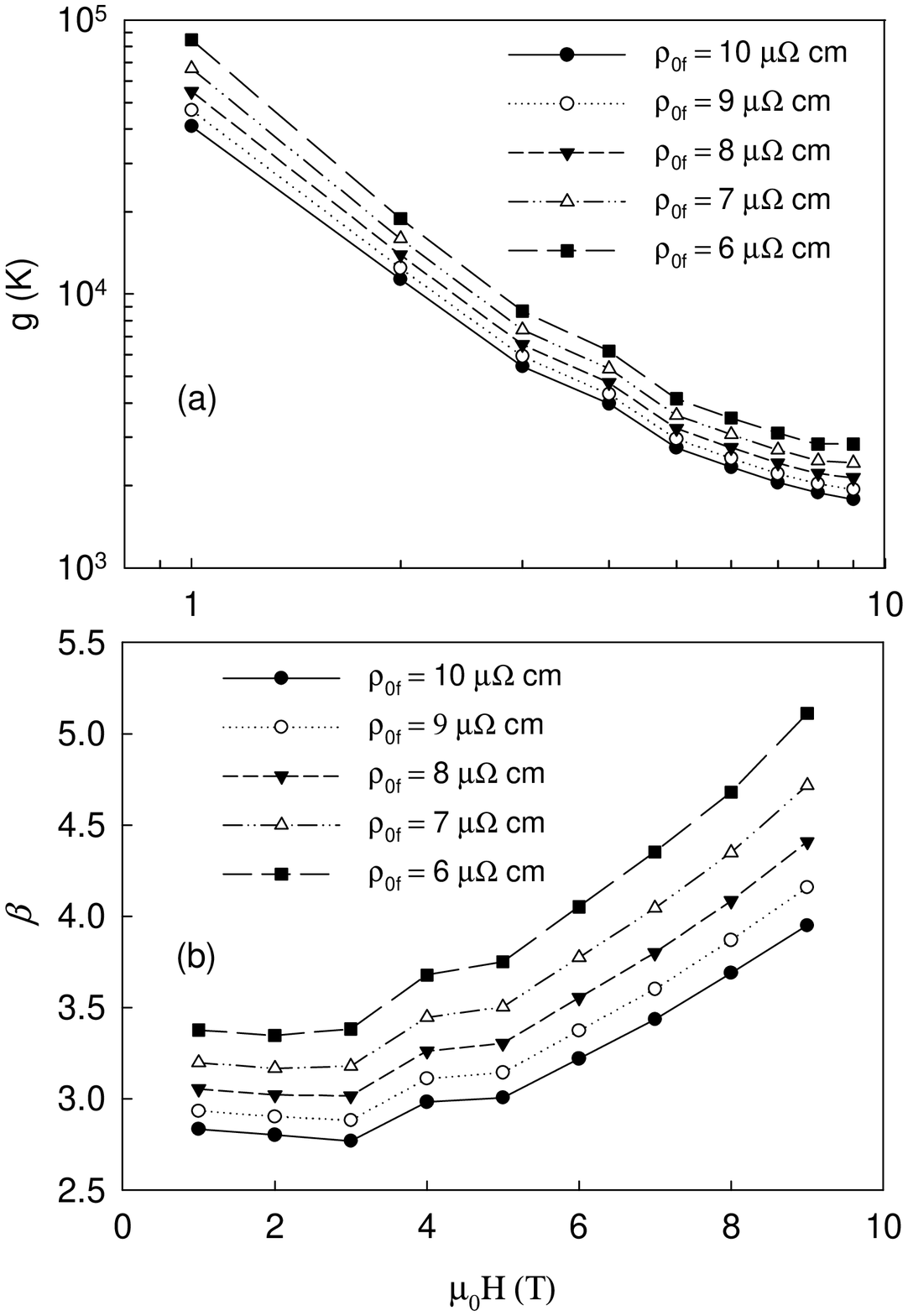} \caption{\label{f3} (a)$g(H)$ and
(b)$\beta(H)$ determined with different $\rho_{0f}$ values.}
\end{figure}

Figures~\ref{f3}(a) and (b) represent $g(H)$ and $\beta(H)$ data
for $\rho_{0f}=6$, 7, 8, 9, 10 $\mu\Omega$ cm, respectively. With
$\rho_{0f}=8$ $\mu\Omega$ cm, we find that $g\sim H^{-1.94}$ and
$\beta\approx 3$ for $1\le \mu_0H\le 3$ T, while  $g\sim
H^{-0.73}$ and $\beta$ approximately linear increases with $H$ for
$5\le \mu_0H\le 9$ T. The characteristic changes of $g$ and
$\beta$ around $\mu_0H=4$ T are probably due to the crossover from
3D to 2D as discussed in Ref.~\cite{Zhang}.

In summary, we suggest that the analysis of activation energies
and definitions of the four different  vortex regions are
incorrect in Ref.~\cite{Kim}, as the TAFF temperature interval
presented in it does not match the interval suggested by Palstra
\textit{et al.} \cite{Palstra}. Using the temperature interval
suggested by Palstra \textit{et al.}, we find that the empirical
activation energy form  suggested by Zhang \textit{et al.}
\cite{Zhang} is in good agreement with the lower resistivity
data.

This work has been financially supported by the National Science
Foundation of China (Grant No. 10174091).


\begin{thebibliography}{}
\bibitem{Kim} W. S. Kim, W. N. Kang, M. S. Kim, and S. I. Lee, Phys. Rev. B \textbf{61}, 11317 (2000).
\bibitem{Zhang} Y. Z. Zhang, Z. Wang, X. F. Lu, H. H. Wen, J. F. de Marneffe, R. Deltour, A. G. M. Jansen, and P. Wyder, Phys. Rev. B \textbf{71},  052502 (2005).
\bibitem{Palstra}T. T. M. Palstra, B. Batlogg, R. B. van Dover, I. F. Schneemeyer, and J. V. Waszczak, Phys. Rev. B \textbf{41}, 6621 (1990).
\bibitem{Geshkenbein}V. B. Geshkenbein, M. V. Feigel\rq man, A. I. Larkin, and V. M. Vinokur, Physica C \textbf{162}, 239 (1989).
\bibitem{Vinokur}V. M. Vinokur, M. V. Feigel'man, V. B. Geshkenbein, and A. I. Larkin, Phys. Rev. Lett. \textbf{65}, 259 (1990);  J. Kierfeld, H. Nordborg, and V. M. Vinokur, Phys. Rev. Lett. \textbf{85}, 4948 (2000).
\bibitem{Blatter} G. Blatter, M. V. Feigel\rq man, V. B. Geshkenbein, A. I. Larkin, and V. M. Vinokur, \rmp \textbf{66}, 1125 (1994).
\bibitem{Brandt}E. H. Brandt,  Rep. Prog. Phys. {\bf 58},
1465 (1995).
\bibitem{Cohen} L. F. Cohen and H. J. Jensen, Rep. Prog. Phys. {\bf 60}, 1581 (1997).

\end{thebibliography}
\end{document}